\RequirePackage{amsmath}
\documentclass[runningheads]{llncs}
\usepackage{dsfont}
\usepackage[arrowdel]{physics} 
\usepackage{multirow}  
\usepackage{xcolor}
\usepackage{amssymb}
\usepackage{lmodern}
\usepackage[T1]{fontenc}
\usepackage{float}
\usepackage{graphics} 
\usepackage{graphicx}
\usepackage{revsymb}
\usepackage{orcidlink}
\hypersetup{hidelinks}

\newtheorem{mytheorem}{Theorem}

\newtheorem{mylemma}[mytheorem]{Lemma}
\newtheorem{mydefinition}[mytheorem]{Definition}
\newtheorem{myproposition}[mytheorem]{Proposition}

\newtheorem{myexample}[mytheorem]{Example}

\usepackage[style=ieee,giveninits=true,sorting=none,maxnames=10,minnames=3]{biblatex}
\usepackage{csquotes}
\newcommand{\1}{\openone} 
\newcommand{\id}{\operatorname{id}}

\newcommand{\ox}{\otimes}

\newcommand{\geqvert}{\rotatebox[origin=c]{270}{$\geq$}}

\newcommand{\CC}{\mathbb{C}}
\newcommand{\EE}{\mathbb{E}}

\newcommand{\cD}{\mathcal{D}}
\newcommand{\cE}{\mathcal{E}}

\newcommand{\cM}{\mathcal{M}}
\newcommand{\cN}{\mathcal{N}}
\newcommand{\cP}{\mathcal{P}}

\newcommand{\cS}{\mathcal{S}}

\newcommand{\cX}{\mathcal{X}}
\newcommand{\cY}{\mathcal{Y}}

\DeclareFieldFormat{journaltitle}{#1\isdot}
\DeclareFieldFormat{publisher}{#1\isdot}
\DeclareFieldFormat{booktitle}{#1\isdot}
\DeclareFieldFormat{title}{#1\isdot}

\newcommand\blfootnote[1]{%
  \begingroup
  \renewcommand\thefootnote{}\footnote{#1}%
  \addtocounter{footnote}{-1}%
  \endgroup
}

\begin{document}
\title{Zero-entropy encoders and simultaneous decoders in identification via quantum channels\textsuperscript{*}\vspace{-0.3cm}}
\titlerunning{Zero-entropy identification and simultaneous decoders\ldots}
%
\author{Pau Colomer\inst{1,4,\orcidlink{0000-0002-8375-8946}} \and
Christian Deppe\inst{2,5,6,\orcidlink{0000-0002-2265-4887}} \and\protect\\
Holger Boche\inst{1,5,6,7,\orcidlink{0000-0002-0126-4521}} \and
Andreas Winter\inst{3,4,\orcidlink{0000-0001-6344-4870}}}
\authorrunning{P. Colomer, C. Deppe, H. Boche, and A. Winter}
%
\institute{Lehrstuhl f\"ur Theoretische Informationstechnik, School of Computation, Information and Technology Technische Universit\"at M\"unchen, Theresienstra{\ss}e 90, 80333 M\"unchen, Germany. \email{\{pau.colomer, boche\}@tum.de}\and 
Institute for Communications Technology, Technische Universit\"at Braunschweig, Schleinitzstraße 22, 38106 Braunschweig, Germany. \email{christian.deppe@tu-bs.de}\and
ICREA \&{} Grup d'Informaci\'o Qu\`antica, Departament de F\'isica, Universitat Aut\`onoma de Barcelona, 08193 Bellaterra, Spain. \email{andreas.winter@uab.cat}\and
Institute for Advanced Study, Technische Universit\"at M\"unchen,\protect\\ Lichtenbergstra{\ss}e 2a, 85748 Garching, Germany\and
6G-life, 6G research hub, Theresienstra{\ss}e 90, 80333 M\"unchen, Germany\and
Munich Quantum Valley (MQV), Leopoldstra{\ss}e 244, 80807 M\"unchen, Germany\and
Munich Center for Quantum Science and Technology,\protect\\ Schellingstra{\ss}e 4, 80799 M\"unchen, Germany} 
\maketitle

\begin{center}
\small \textsc{In memory of Ning Cai}
\end{center}

\begin{abstract}
Motivated by deterministic identification via classical channels, where the encoder is not allowed to use randomization, we revisit the problem of identification via quantum channels but now with the additional restriction that the message encoding must use pure quantum states, rather than general mixed states. Together with the previously considered distinction between simultaneous and general decoders, this suggests a two-dimensional spectrum of different identification capacities, whose behaviour could a priori be very different. 

We demonstrate two new results as our main findings: first, we show that all four combinations (pure/mixed encoder, simultaneous/general decoder) have a double-exponentially growing code size, and that indeed the corresponding identification capacities are lower bounded by the classical transmission capacity for a general quantum channel, which is given by the Holevo-Schumacher-Westmoreland Theorem. Secondly, we show that the simultaneous identification capacity of a quantum channel equals the simultaneous identification capacity with pure state encodings, thus leaving three linearly ordered identification capacities. By considering some simple examples, we finally show that these three are all different: general identification capacity can be larger than pure-state-encoded identification capacity, which in turn can be larger than pure-state-encoded simultaneous identification capacity.
\vspace{-0.2cm}

\end{abstract}

\vspace{-0.66cm}
\blfootnote{{\textsuperscript{*}A preliminary version of this work was presented at the 2024 ICC
\cite{CDBW:ID-0-sim:ICC}.}}

\section{Introduction}
The classical communication model described by Shannon \cite{Shannon:TheoryCommunication} revolves around the concept of a noisy channel $W:\cX\rightarrow\cY$ between an input alphabet $\cX$ and an output alphabet $\cY$. When these alphabets are discrete or even finite, the channel can be described as a stochastic matrix of transition probabilities $W(y|x)$ from an input letter $x\in\cX$ to an output $y\in\cY$. In block length $n\in \mathds{N}$, we have $n$-letter sequences (words) $x^n=x_1x_2\dots x_n\in \cX^n$ and $y^n=y_1y_2\dots y_n\in \cY^n$, whose transition probabilities are given by
$W^n(y^n|x^n)=\prod_{t=1}^n W(y_t|x_t)$.

Shannon proved reliable transmission of messages can be achieved through noisy channels. This means that a sender can transmit messages to a receiver with an error probability that decreases as the block length increases, reaching zero in the limit of $n\rightarrow\infty$. Moreover, the number $M$ of messages that can be transmitted reliably increases exponentially with the block length, $M=2^{nR}$, where $R$ is the rate of a transmission code, the ratio between the length of the messages in bits ($\log M$) and the block length ($n$). 

\begin{mydefinition}
\label{def:ctrans}
An $(n,M,\lambda)$-\emph{transmission code} over $n$ uses of the memoryless channel $W$ is a family of pairs $((u_m,\cD_m):m\in[M]=\{1,\dots,M\})$ with $u_m\in\cX^n$ code words and $\cD_m\subset\cY^n$ pairwise disjoint subsets of the output words (for all $m\neq m'\in[M]$, $\cD_m\cap\cD_{m'}=\emptyset$) such that the error probability is bounded by $\lambda$:
\(
    W^n(\cD_m|u_m)\geq 1-\lambda 
\)
for all $m\in[M]$.
\end{mydefinition}
Denoting by $M(n,\lambda)$ the maximum number of messages of an $(n,M,\lambda)$-code, we can define the \emph{capacity} $C(W)$ of the noisy channel as the maximum possible rate for asymptotically faithful transmission:
\begin{equation}
  \label{eq:Trans_capacity_def}
  C(W) := \inf_{\lambda>0}\liminf_{n\rightarrow\infty} \frac{1}{n}\log M(n,\lambda).
\end{equation}

Here and elsewhere in this article $\log$ and $\exp$ are to base 2 by default. 
For later use, let us define some information theory notations. First, let $\cP(\cX)$ be the set of probability distributions on $\cX$; then, the mutual information is given by $I(P;W)=H(PW)-H(W|P)$, where $PW = \sum_x P(x)W_x \in \cP(\cY)$, and with the entropy $H(Q)=-\sum_y Q(y)\log Q(y)$ and the conditional entropy $H(W|P)=-\sum_x P(x) H(W(\cdot|x))$.

\begin{mytheorem}[\cite{Shannon:TheoryCommunication,Wolfowitz:converse}]\label{thm:Shannon_coding}
For all $\lambda\in (0;1)$, we have the capacity of communication and the strong converse given by
\[
  C(W) = \lim_{n\rightarrow\infty}\frac{1}{n}\log M(n,\lambda)
       = \max_{P\in\cP(\cX)} I(P;W).
\]
\end{mytheorem}

\begin{figure}
    \centering
    \includegraphics[width=0.9\linewidth]{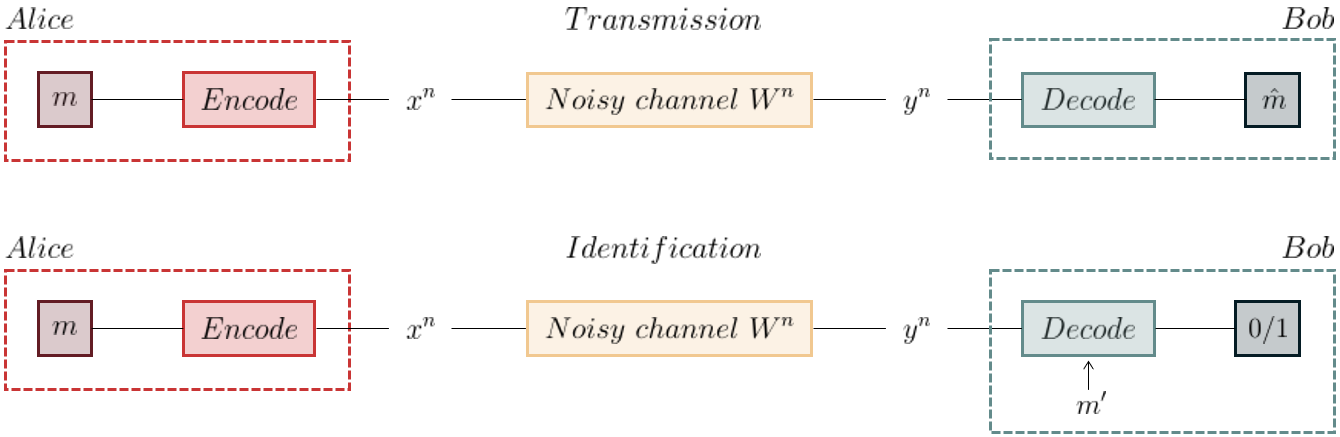}
    \caption{\small{Let Alice encode a message $m$ from a set $\cM=\{1,\dots,M\}$ into a code word of length $n$ and send it through a discrete memoryless channel (DMC) described by the stochastic matrix $W^n$. In the usual transmission scheme (above), when Bob receives $y^n$ he can decode the message, aiming to recover some $\hat{m}\approx m$. In an identification scheme (below), he instead chooses any message $m'\in\cM$ and checks whether it is equal to $m$ with a particular hypothesis testing decoder, obtaining a binary answer.}}
    \label{fig:Trans_VS_ID}
\end{figure}

The concept of \emph{identification} of messages via channels was first introduced by Ahlswede and Dueck in their seminal paper \cite{AD:ID_ViaChannels}, which drew inspiration from the earlier work of JaJa \cite{Ja:ID_easier} and Yao \cite{Yao:complexity}. In this task, instead of recovering an original message, the primary objective of the receiver is to determine whether it matches a specific message that he has in mind. The output in identification merely consists of a single bit indicating the match or mismatch of the sent message and the one the receiver wants to identify (see Figure \ref{fig:Trans_VS_ID}).

Ahlswede and Dueck proved that identification codes (\emph{ID codes}) can achieve double exponential growth of the message set as a function of the block length $n$. This represents a dramatic improvement over transmission codes, which exhibit only exponential codes in $n$. We can identify an exponentially larger number $N\sim2^{2^{nR}}$ of messages than we can transmit. The main insight to prove this surprising achievability improvement comes from utilizing randomness on the encoder (inspired by \cite{Yao:complexity}). Instead of a code word $u_m\in\cX^n$ for each message $m\in[M]$ as we had in the transmission scenario, we now have a probability distribution $P_j=P(\cdot|j)\in\cP(\cX^{n})$ for each message $j\in[N]$:

\begin{mydefinition}\label{def:clasical_ID}
A (randomized) $(n,N,\lambda_1,\lambda_2)$\emph{-ID code} is a family $\{(P_j,\cE_j):j\in[N]\}$ with probability distributions $P_j\in\cP(\cX^{n})$ and decoding sets $\cE_j\subset\cY^n$, such that, reusing the notation $P_j W^n=\sum_{x^n}P(x^n|j)W_{x^n}$, for all $j\neq k\in [N]$ it holds
\begin{align}
    (P_jW^n)(\cE_j) &\geq 1-\lambda_1, \label{eq:clas_1error}\\
    (P_jW^n)(\cE_k) &\leq \lambda_2. \label{eq:clas_2error}
\end{align}
\end{mydefinition}

Notice the formal differences between identification and transmission codes. In randomized identification, we have probability distributions on the input and we do not require disjointness of the output decoding sets. These two factors yield the appearance of two possible errors termed \emph{first} and \emph{second kind}, following the standard terminology in hypothesis testing, in contrast to transmission where we only have a single maximum error probability of incorrectly decoding $\lambda$. Here, $\lambda_1$ is the probability of a missed identification [Equation \eqref{eq:clas_1error}]. This happens when the message Alice sends is the same as the one Bob wants to identify but due to the noise of the protocol, the hypothesis test has a negative outcome. Likewise, $\lambda_2$ is the probability of incorrect identification [Equation \eqref{eq:clas_2error}]: the messages sent and tested are different, but the outcome on Bob's side is positive.

The main idea in Ahlswede and Dueck's achievability proof is the observation that given any finite set, there is an exponential number of subsets of the ground set that have pairwise small overlaps:

\begin{myproposition}[\cite{AD:ID_ViaChannels}]\label{prop:A&D}
    Let $\cM$ be a set of cardinality $M$, $\lambda>0$ and $\epsilon$ such that $\lambda\log\left(\frac{1}{\epsilon}-1\right)>2$. Then there exist $N\geq2^{\lfloor{\epsilon M}\rfloor}/M$ subsets $\cM_j\subset\cM$ of cardinality $\lfloor
    \epsilon M\rfloor$, such that 
    \[
    \forall j\neq k\quad |\cM_j\cap\cM_k|\leq\lambda\lfloor{\epsilon M} \rfloor.
    \]
\end{myproposition}
They prove then that starting from an $(n,N,\lambda)$-transmission code reliable identification can be achieved by encoding an exponentially larger number $N\sim 2^{2^{nR}}$ of messages into uniform distributions $P_j$ over the subsets $\cM_j\subset\cM$ of the transmission code words according to Proposition \ref{prop:A&D}. Thus, they defined the double exponential identification capacity over a channel $W$ as
\begin{equation}\label{eq:C_ID_definition}
    C_\text{ID}(W):=\inf_{\lambda_1,\lambda_2>0}\liminf_{n\rightarrow\infty}\frac1n \log\log N(n,\lambda_1,\lambda_2),
\end{equation}
where $N(n,\lambda_1,\lambda_2)$ denotes the maximum number of messages $N$ such that an $(n,N,\lambda_1,\lambda_2)$-ID code exists. By the above concatenation of a transmission code with Proposition \ref{prop:A&D}, they established that the double exponential capacity of identification via a discrete memoryless channel is at least as large as the Shannon (single) exponential transmission capacity of the channel (in Theorem \ref{thm:Shannon_coding}). They also showed optimality in a ``soft'' sense, while Han and Verdú \cite{HanVerdu:ID} were the first to prove the full (strong) converse:

\begin{mytheorem}[\cite{AD:ID_ViaChannels,HanVerdu:ID}]
\label{thm:Classical_ID_capacity}
The double exponential ID capacity of a channel $W$ equals Shannon's (single) exponential transmission capacity, and the strong converse holds: for any $\lambda_1,\lambda_2>0$, $\lambda_1+\lambda_2<1$,
\[
    C_{\text{ID}}(W) 
     = \lim_{n\rightarrow\infty} \frac1n \log\log N(n,\lambda_1,\lambda_2)
     = C(W).
\]
\end{mytheorem}
(Note here that we exclude the case $\lambda_1+\lambda_2\geq 1$ as trivial: indeed, any number of messages can be encoded in this regime by simply making all $P_j$ arbitrary and equal.)

We have seen that the general ID codes from Definition \ref{def:clasical_ID} use a randomized encoder, meaning that the input string is not deterministically defined (it could be any $x^n$ with $P(x^n|j)\neq0)$. But what happens if we impose a deterministic encoding that uses code words $u_j\in\cX^n$, like in the transmission scenario, instead of probability distributions? (In the definition, this means that the $P_j={\delta_{u_j}}$ are point masses.) This question interested researchers from the beginning because deterministic codes are usually easier to implement, construct and simulate \cite{BAAS:Easy_DI_simulation,DI_explicit_construction} and offer reliable single-block performance \cite{AD:ID_ViaChannels}. It was observed that the deterministic approach leads to poorer results in terms of the scaling of the code in the block length \cite{AD:ID_ViaChannels,AC:DI,SPBD:DI_power}. Indeed, it was noted in these works and formally proved in \cite{SPBD:DI_power} that identification over discrete memoryless channels can only lead to exponential scaling like in Shannon's paradigm. For other channels with continuous input and/or output alphabets, it was found that deterministic codes are governed by a slightly super-exponential scaling in block length $N\sim2^{Rn\log n}$ \cite{DI-fading,DI-poisson}.

In the present paper, we explore in depth what happens to identification over quantum channels, developed by L\"ober \cite{loeber:PhDThesis} and since \cite{AW:StrongConverse,Winter:QCidentification,Winter:QC-ID-randomness,HaydenWinter:QC-ID,Winter:Review,BDW:ID-compound,TSW:soft-covering}, if we restrict the encoder to not being allowed to randomize but otherwise full access to quantum effects; we call this model \emph{zero-entropy encoders}.
In the next Section \ref{sec:cID} we review identification of classical messages over quantum channels, including the fundamentally quantum distinction between simultaneous and non-simultaneous (general) decoders, and we define the model of zero-entropy encoders. 
In Section \ref{sec:pure-encoders}, we show the first of our new results, namely that zero-entropy encoders (with, additionally, simultaneous decoders) still achieve double exponential message growth, and that the double-exponential rate is indeed lower bounded by the (single exponential) transmission capacity. On the way, we show that transmission codes may with little loss of generality employ orthogonal states at the encoder. 
In Section \ref{sec:C_train}, we move to the second main result, namely that the simultaneous ID capacity is attained with zero-entropy encoders, by modifying the encoder of a given simultaneous ID code while not touching the decoder. This sheds new light on the work of L\"ober and the insights from Section \ref{sec:pure-encoders}. 
In Section \ref{sec:examples}, we collect all new insights to observe that the old and new ID capacities over quantum channels arising from various combinations of constraints on encoders and decoders result in a linearly ordered chain of quantities, which we separate by a simple family of example channels. 
We conclude in Section \ref{sec:conclusions}.

\section{Transmission and identification via quantum channels}
\label{sec:cID}
All the concepts defined up to now can be generalised to quantum channels quite naturally. Let us start with some preliminaries to fix notions and notation; see \cite{holevo:stat-structure-book,Wilde-book,QCN} for a comprehensive treatment of the quantum formalism in the present context. 

Quantum systems are described by complex Hilbert spaces $A$, $B$, etc, whose dimension is denoted $|A|$, $|B|$, etc. (In the present paper, all Hilbert space dimensions will be finite.) 
Our basic object of interest is a quantum channel $\cN$, which is a completely positive and trace-preserving (cptp) linear map between the states of two finite-dimensional complex Hilbert spaces $A$ and $B$, $\cN:A\rightarrow B$. The cptp condition is necessary and sufficient to ensure that any quantum state $\rho\in\cS(A)$ in the set of quantum states in the Hilbert space $A$ (i.e.~a positive semidefinite density operator $\rho\geq0$ with $\Tr\rho=1$), is mapped to another quantum state $\cN(\rho)\in\cS(B)$ on the output, and that the same holds for all $\cN\ox\id_C$ with arbitrary auxiliary quantum systems $C$. 

To extract information from a quantum state one has to measure it; quantum measurements are described by positive operator-valued measures (POVMs). In the discrete case of interest here, a POVM is a collection $(K_t)_{t\in\mathcal{K}}$ of positive semidefinite operators, i.e.~$K_t\geq0$, that sum to the identity, $\sum_{t\in\mathcal{K}} K_t=\1$. In the present paper, we find it convenient to relax the definition of POVMs to sub-POVMs, where the sum of the elements only has to be smaller than the identity: $\sum_{t\in\mathcal{K}} K_t\leq\1$. The probability of obtaining the outcome $t$ when measuring $\rho\in\cS(A)$ with the POVM $(K_t)_{t\in\mathcal{K}}$ on $A$ is given by the Born rule: $\Pr\{t|\rho\} = \Tr \rho K_t \in [0;1]$. The probability is $0$ if and only if the quantum state is orthogonal to the POVM element (equivalently: if the support of $\rho$ is contained in the kernel of $K_t)$, and $1$ if and only if the support of $\rho$ is contained in the $1$-eigenspace of $K_t$. 

Measurements give rise to statistical distance measures on the set of quantum states, of which we here need the \emph{trace distance} and the \emph{fidelity}. The trace distance of two density matrices $\rho$ and $\sigma$ is 
\[
  \frac{1}{2}\|\rho-\sigma\|_1 
    :=\frac12\Tr\left|\rho-\sigma\right|
    = \frac12\Tr\sqrt{(\rho-\sigma)^2}.
\]
In particular, the trace distance between two diagonal density matrices (which correspond to classical probability distributions) is the total variation distance between the distributions.
The fidelity of quantum states on the other hand is 
\[
  F(\rho,\sigma)
   := \|\sqrt{\rho}\sqrt{\sigma}\|_1
   = \Tr\sqrt{\sqrt{\rho}\sigma\sqrt{\rho}} 
   = \min_{(F_i)} \sum_i\sqrt{(\Tr\rho F_i)(\Tr\sigma F_i)},
\]
where $F_i$ are the elements of the POVM $(F_i)$, i.e.~$F_i\geq 0$ and $\sum_i F_i=\1$. The fidelity and the trace distance of quantum states are related to each other through the Fuchs-van-de-Graaf inequalities \cite{FVdG:ineq}:
\begin{equation*}
  1-F(\rho,\sigma)\leq\frac{1}{2}\|\rho-\sigma\|_1\leq\sqrt{1-F(\rho,\sigma)^2}.
\end{equation*}

\begin{mydefinition}
\label{def:qtrans}
An $(n,M,\lambda)$\emph{-code} for classical communication over a quantum channel $\cN:A\rightarrow B$ is a collection of quantum states $\pi_m$ on $A^n=A^{\otimes n}$ and POVM elements $D_m$ acting on $B^n=B^{\otimes n}$ ($m\in\cM$ with $|\cM|=M$), such that
\begin{equation}
  \label{eq:Tr_error}
  \forall m\in\cM\quad 
  \Tr \cN^{\otimes n}(\pi_m) D_m \geq 1-\lambda.
\end{equation}
\end{mydefinition}
This definition is completely analogous to the classical one (Definition \ref{def:ctrans}), but using the quantum formalism on the input, the output, and the probability of error, allowing us to maintain the definitions for the communication rate $R:=\frac1n\log M$, the maximum number of messages $M(n,\lambda)$ in an $(n,M,\lambda)$-code and the (now quantum) channel capacity for classical transmission of messages as in Equation \eqref{eq:Trans_capacity_def}. 

Holevo, Schumacher and Westmoreland \cite{holevo:capacity,SW:capacity} determined the classical capacity of a general channel $\cN$, and also the capacity under the added restriction that the encodings only use separable states. Let us denote the maximum number of messages that a code with such restriction can achieve by $M_\text{sep}(n,\lambda)$ and the corresponding restricted capacity as
\[
  C_\text{sep}(\cN):=\inf_{\lambda>0}\liminf_{n\rightarrow\infty}\frac1n\log M_\text{sep}(n,\lambda).
\]
The strong converse under this restriction was proved in \cite{ON:Strong_converse,Winter:Strong_converse}. For general encodings, the capacity is known but in general only 
a weak converse \cite{Holevo:capacity0,holevo:capacity,SW:capacity,winter:PhDThesis}. 

\begin{mytheorem}[\cite{holevo:capacity, SW:capacity,ON:Strong_converse,Winter:Strong_converse}]
\label{thm:HSW+}
For all $\lambda\in(0,1)$, we have the coding theorem and strong converse for communicating with separable states via a channel $\cN$, given by
\[
  C_\text{sep}(\cN)
  =\lim_{n\rightarrow\infty}\frac1n\log M_\text{sep}(n,\lambda)
  =\chi(\cN),
\]
with the \emph{Holevo capacity}
\[
  \chi(\cN) = \max_{\{p_x,\pi_x\}} \left[H\left(\sum_x p_x\cN(\pi_x)\right) - \sum_x p_x H\left(\cN(\pi_x)\right)\right], 
\]
where $H(\rho)=-\Tr\rho\log\rho$ is the von Neumann entropy. 

For general encodings, the capacity and the weak converse are given by the regularization of the Holevo capacity:
\[
  C(\cN)
  =\inf_{\lambda>0} \limsup_{n\rightarrow\infty} \frac{1}{n}M(n,\lambda)
  =\lim_{n\rightarrow\infty}\frac{1}{n}\chi(\cN^{\otimes n}).
\]
\end{mytheorem}
Notice that Theorem \ref{thm:HSW+} generalizes the classical result (Theorem \ref{thm:Shannon_coding}) because classical strings are separable and the Holevo capacity of a classical channel $W$ equals the Shannon transmission capacity: 
\[
  \chi(W) = \max_{P\in\cP(\cX)}I(P;W) 
          = C(W).
\]

In the same way, we have generalized the transmission of classical messages from classical to quantum channels, L\"ober did the same for the identification task \cite{loeber:PhDThesis}. Later on, this problem was further expanded by Ahlswede and Winter \cite{AW:StrongConverse} and in subsequent work \cite{Winter:QCidentification,Winter:QC-ID-randomness,HaydenWinter:QC-ID,Winter:Review,BDW:ID-compound,TSW:soft-covering}:

\begin{mydefinition}
\label{def:qID_code}
An $(n,N,\lambda_1,\lambda_2)$\emph{-ID~code} is a collection of $N$ quantum states $\rho_j$ on $A^n$ and effect operators $0\leq E_j\leq \1$ acting on $B^n$ ($j\in[N]$), such that
\begin{equation}\label{eq:ID_errors}
    \begin{split}
        \forall j\quad &\Tr \cN^{\otimes n}(\rho_j)E_j \geq 1-\lambda_1,\\
        \forall j\neq k \quad &\Tr \cN^{\otimes n}(\rho_j)E_k \leq \lambda_2.
    \end{split}
\end{equation}
\end{mydefinition}

Instead of having a POVM $(D_m)_{m\in\cM}$ on the output system as in the transmission scenario, in identification we apply a quantum binary hypothesis test $(E_j,\1-E_j)$, so the different positive operators $E_j$ do not have to be elements of a POVM. In fact, the POVMs $(E_j,\1-E_j)$ need not even be simultaneously measurable, unlike the classical case, due to the complementarity of quantum observables, manifest in the non-commutativity of their operators. However, if the tests $(E_j,\1-E_j)$ do indeed descend from a single POVM by coarse-graining, we say that the code is simultaneous:

\begin{mydefinition}
\label{def:sim-ID-code}
An ID code is called \emph{simultaneous} if the binary POVMs $(E_j,\1-E_j)$ are coexistent. In other words, if there exists a POVM $(D_m)_{m\in\cM}$ such that for all $j$ there are subsets $\cM_j\subset\cM$ satisfying $E_j=\sum_{m\in \cM_j} D_m$.
\end{mydefinition}

The maximum $N$ such that an $(n,\lambda_1,\lambda_2)$-ID code exists is denoted here as $N(n,\lambda_1,\lambda_2)$ and the identification capacity is defined like in the classical case, see Equation \eqref{eq:C_ID_definition}. If the code is restricted to be simultaneous, we similarly define $N_\text{sim}(n,\lambda_1,\lambda_2)$ and a double exponential simultaneous ID capacity over the channel $\cN$:
\begin{equation}\label{eq:C_id^sim_def}
C_{\text{ID}}^{\text{sim}}(\cN) := \inf_{\lambda_1,\lambda_2>0} \liminf_{n\rightarrow\infty} \frac1n \log\log N_{\text{sim}}(n,\lambda_1,\lambda_2).
\end{equation}

To prove the double exponential achievability of the ID code, L\"ober used the same idea as in the classical setting. We reproduce his proof here as it can serve as intuition for our posterior calculations. We have to start from an $(n,M,\lambda)$-code for transmission of classical messages over quantum channels $\{(\pi_m,D_m):1,\dots,M\}$ (see Definition \ref{def:qtrans}) with $M\geq2^{(C-\zeta)n}$ for some $\zeta>0$. Proposition \ref{prop:A&D} tells us that it is possible to find for $n\gg1$ a number 
\[N\geq2^{\lfloor\epsilon M\rfloor}/M \geq 2^{\lfloor\epsilon 2^{(C-\zeta)n}-n\rfloor}\]
of subsets of the transmission states with pairwise small overlaps. In the classical scenario, we created uniform distributions of the transmission code words (classical strings) in each subset; now that the code words (and therefore the subset elements) are quantum states, we write instead the uniform mixture of the code states in each subset
\begin{equation}\label{eq:LoeberMixture}
    \rho_j = \frac{1}{\lfloor\epsilon M\rfloor} \sum_{m\in\cM_j}\pi_m.
\end{equation} 
Finally, defining the decoding operators $E_j=\sum_{m\in\cM_j} D_m$ for all $j\in[N]$, we observe that the errors can be bounded for all $j\neq k$: 
\begin{equation}
\label{eq:Lober_ErrorAnalysis}
\begin{split}
    \Tr \cN^{\otimes n}(\rho_j)E_j 
    &= \Tr\left[\cN^{\otimes n}\left( \frac{1}{\lfloor\epsilon M\rfloor} \sum_{m\in\cM_j}\pi_m\right)\!\sum_{m'\in\cM_j}\!\!\! D_{m'}\right]\\
    &\geq \frac{1}{\lfloor\epsilon M\rfloor} \sum_{m\in\cM_j}\!\!\Tr \cN^{\otimes n}(\pi_m)D_m \geq 1-\lambda,\\        
    \Tr \cN^{\otimes n}(\rho_j)E_k 
    &=\frac{1}{\lfloor\epsilon M\rfloor}\! \sum_{m\in\cM_j} \!\!\left[\!\!\!\!\sum_{\phantom{=}m'\in\cM_j\cap\cM_k}\!\!\!\!\!\!\!\!\!\!\!\Tr\cN^{\otimes n}(\pi_m)D_{m'}+\!\!\!\!\!\!\!\!\!\!\sum_{m'\in\cM_j\setminus\cM_k}\!\!\!\!\!\!\!\!\!\Tr\cN^{\otimes n}(\pi_m)D_{m'}\right] \\
    &\leq \frac{1}{\lfloor\epsilon M\rfloor} \left[\sum_{m'\in\cM_j\cap\cM_k}\!\!\!1+\!\!\!\sum_{m'\in\cM_j\setminus\cM_k}\!\!\!\lambda\right]\\
    &=\frac{1}{\lfloor\epsilon M\rfloor}\cdot\lambda\lfloor\epsilon M\rfloor+\frac{1}{\lfloor\epsilon M\rfloor}\cdot\lfloor\epsilon M\rfloor\cdot\lambda=2\lambda.
    \end{split}
\end{equation}
That is, with $\lambda_1=\lambda$ and $\lambda_2=2\lambda$ this construction achieves an $(n,N,\lambda_1,\lambda_2)$-ID code with $N$ double exponential in the block length. Furthermore, notice that the code created with this scheme is simultaneous. 

In conclusion, any transmission code over the quantum channel $\cN$ with error $\lambda$, yields an exponentially larger simultaneous ID code with errors of first and second kind $\lambda_1=\lambda$ and $\lambda_2=2\lambda$. Therefore, it is clear that the double exponential capacity of simultaneous identification in Equation \eqref{eq:C_id^sim_def} can be lower-bounded by the capacity of transmission. It is not known whether these two capacities are equal in general: neither do we have an example of a channel where they are different, nor a converse that proves the equality.

Also, the simultaneous identification capacity is clearly a lower bound on the unrestricted identification capacity $C_\text{ID}$, as we are imposing a restricting condition on the hypothesis test operators. We know examples where the simultaneous ID capacity and the unrestricted ID capacity are different ($2=C_\text{ID}(\id_2) > C_\text{ID}^\text{sim}(\id_2)=1$ \cite{Winter:QCidentification,TSW:soft-covering}, see Example \ref{example:id}, and especially Proposition \ref{prop:upperbound_sizeN} for further intuition), as well as examples of channels for which they are the same (for instance all cq-channels \cite{AW:StrongConverse}). All this knowledge results in the following chain of inequalities:
\begin{equation}
  \label{eq:old_train}
  C_\text{ID}(\cN) \geq C_\text{ID}^\text{sim}(\cN) 
  \geq C(\cN),
\end{equation}
with the open problem regarding the possible collapse of the second inequality in general.

\medskip
At this point, we can formulate the problem to be addressed in the present paper. We would like to define a quantum channel version of deterministic identification, which we take to be characterised by not allowing randomisation at the encoder. Just as in the classical case of Ahlswede and Dueck \cite{AD:ID_ViaChannels}, L\"ober's code construction results in highly mixed states as message encodings. On the other hand, the closest analogue of a point mass in quantum mechanics is a pure state, the least random of density matrices, and indeed characterised by having von Neumann entropy zero. 
Pure state encodings for identification have actually been considered early on in quantum information theory under the name of \emph{quantum fingerprinting}, for the noiseless channel, and it had been shown that the message set grows double exponentially in the block length \cite{BCW:fingerprinting}. 


However, it turns out that things are not that simple. The classical deterministic identification code words are different perfectly distinguishable sequences. Quantum states are only perfectly distinguishable if they are orthogonal. Thus, if were to maintain this feature in our quantum construction, we end up with pure orthogonal states. However, pure orthogonal states are essentially classical. If we want to construct a code that takes advantage of the quantum features of the channel input system, it is necessary to allow for some overlap between the states (coherences). 

\begin{mydefinition}
\label{def:zero-entropy-ID}
An $(n,N,\lambda_1,\lambda_2)$-ID code $\{(\rho_j,E_j):j\in[N]\}$ for the memoryless channel $\cN$ is said to have \emph{zero-entropy encoder} is all $\rho_j\in\cS(A^n)$ are pure states. 

Define $N_0(n,\lambda_1,\lambda_2)$ as the maximum number of messages in an ID code with zero-entropy encoder, and $N_{\text{sim},0}(n,\lambda_1,\lambda_2)$ the same if the ID code is additionally required to be simultaneous.
\end{mydefinition}

As before, from these we can obtain the asymptotic zero-entropy ID capacities $C_{\text{ID}}^0(\cN)$ and $C_{\text{ID}}^{\text{sim},0}(\cN)$, defined in the double exponential scale:
\begin{equation}\begin{split}
   C_\text{ID}^{0} 
   &:=\inf_{\lambda_1,\lambda_2>0} \liminf_{n\rightarrow\infty} \frac{1}{n}\log\log N_{0}(n,\lambda_1,\lambda_2), \\
   C_\text{ID}^{\text{sim},0}
   &:=\inf_{\lambda_1,\lambda_2>0} \liminf_{n\rightarrow\infty} \frac{1}{n}\log\log N_{\text{sim},0}(n,\lambda_1,\lambda_2).
\end{split}\end{equation}
These capacities can only be non-zero if the code size scales double exponentially with the block length, a feature that cannot be achieved by deterministic identification via classical channels. The existence of quantum fingerprinting for the noiseless channel shows that quantum systems behave differently. 
Finding positive values for these capacities would mean that zero-entropy encoders can achieve better performances than what we expect for an analogue to deterministic identification over quantum channels. 

As encoding only in pure states is a restriction, it is obvious that $C_\text{ID}(\cN) \geq C_\text{ID}^0(\cN)$. With the same considerations and taking into account the distinction between simultaneous and non-simultaneous identification \cite{loeber:PhDThesis,AW:StrongConverse}, we can construct a two-dimensional spectrum of different identification capacities, whose behaviour could in principle be completely different:
\begin{eqnarray}
\label{eq:2D}
  C_{\text{ID}}(\cN) & \geq & C_{\text{ID}}^{\text{sim}}(\cN) \nonumber\\
  \geqvert\phantom{=:} & & \phantom{==}\geqvert \\
  C_{\text{ID}}^0(\cN) & \geq & C_{\text{ID}}^{\text{sim},0}(\cN) \nonumber
\end{eqnarray}

All the inequalities in \eqref{eq:2D} can be understood from the reasoning that we are adding extra restrictions to our codes. Notice that this logic a priori does not give us any information about the relation between $C_\text{ID}^0(\cN)$ and $C_\text{ID}^\text{sim}(\cN)$. 

\section{Zero-entropy encoders (with simultaneous decoders) achieve double exponential identification codes}
\label{sec:pure-encoders}
Following L\"ober, concatenating an $(n,\lambda)$-transmission code for a channel with Ahlswede/Dueck's Proposition \ref{prop:A&D}, we will get a simultaneous ID code with double exponentially growing messages $N$. Our idea is to replace the uniform mixtures over subsets $\cM_i\subset\cM$ with \emph{superpositions}, while keeping L\"ober's decoder, with essentially the same code parameters. For this to work, we require a channel code with pure and mutually orthogonal code states, so as to have a controllable normalisation constant in the superposition. The following result proves that we can construct such a code from any general transmission code without affecting much the maximum probability of error and the size of the code. We formulate it in a one-shot setting without assumptions on memorylessness. 

\begin{mytheorem}\label{th:PureOrthogonalCode}
Given a transmission code $\{(\pi_m,D_m):m\in\cM\}$ over a quantum channel $\cN$ with $M=|\cM|$ messages and average error probability $\varepsilon$, meaning 
\[
  \frac{1}{M}\sum_{m\in\cM} \Tr\cN(\pi_m)D_m \geq 1-\varepsilon.
\]
Then, one can construct a code $\{(\varphi_m,D_m):m\in\cM'\subset\cM\}$ such that the $M'=|\cM'|$ code words $\varphi_m$ are pure and mutually orthogonal states, where $M'\geq (1-\varepsilon)^2M/2$, and having maximum error probability $\delta \leq 2\sqrt{5\varepsilon}/(1-\varepsilon)^2$:
\[
  \forall m\in\cM'\quad 
  \Tr\cN(\varphi_m)D_m \geq 1-\delta.
\]
\end{mytheorem}

\begin{proof}
Without loss of generality, $\cM=[M]$.
Our $\{(\pi_m,D_m):m\in[M]\}$ transmission code has average probability of error bounded by $\varepsilon$:
\[
    \bar{P}_e = \frac{1}{M}\sum_{m=1}^M \left(1-\Tr \cN(\pi_m)D_m\right) 
    = \frac{1}{M}\sum_{m=1}^M \left(1-\Tr \pi_m E_m\right)
    \leq \varepsilon,
\]
where we have defined the POVM $E_m=\cN^*(D_m)$, with the adjoint map $\cN^*$ of the given channel, which is cpup (completely positive and unit-preserving) and in particular maps (sub-)POVMs to (sub-)POVMs. 
As any state $\pi_m$ can be written as a convex combination of pure states, there necessarily exist pure states $\psi_m$ such that
\[
  \frac{1}{M}\sum_{m=1}^M \left(1-\Tr \psi_m E_m\right) \leq \varepsilon.
\]

Now we invoke the following result:
\begin{mylemma}[{Barnum/Knill~\cite{BK:PGM}}]
\label{lemma:BK}
Given quantum states $\rho_m$, 
if there exists a POVM $(E_m:m\in[M])$ such that $\frac{1}{M}\sum_{m=1}^M \Tr \rho_m E_m \geq 1-\varepsilon$, then the pretty good measurement (PGM) $(F_m:m\in[M])$ with elements $F_m=S^{-1/2}\rho_m S^{-1/2}$, where $S=\sum_{m}\rho_m$, achieves
\[
  \frac{1}{M}\sum_{m=1}^M \Tr F_m\rho_m \geq (1-\varepsilon)^2.
\]
\end{mylemma}
Notice that $\sum_{m=1}^MF_m=\sum_{m=1}^MS^{-\frac{1}{2}}\psi_mS^{-\frac{1}{2}}=S^{-\frac{1}{2}}SS^{-\frac{1}{2}}=\Pi_L$ is a projector onto the support of S of dimension $L$. Furthermore,
\[
  (1-\varepsilon)^2\leq\frac{1}{M}\sum_{m=1}^M\Tr\psi_mF_m\leq\frac{1}{M}\sum_{m=1}^M\Tr F_m=\frac{L}{M},
\]
so $L\geq M(1-\varepsilon)^2$. In other words, we can find ``many'' ($L$) linearly independent elements among the pure states $\psi_m$. Without loss of generality, we may assume them to be the first $L$ of the initial $M$ states. The code consisting of these specific linearly independent states and the corresponding unmodified decoder has an average probability of error given by
\begin{equation}\label{eq:li_code}
    \frac{1}{L}\sum_{m=1}^L \left(1-\Tr \psi_m E_m \right)
    \leq\frac{M}{L}\frac{1}{M}\sum_{m=1}^M \left(1-\Tr \psi_m E_m\right)
    \leq\frac{\varepsilon}{(1-\varepsilon)^2}.
\end{equation}
From \cite{Holevo:PGM} we know that the PGM elements defined from linearly independent states are orthogonal. So we construct again a PGM, but now only with the linearly independent elements. I.e., defining $T=\sum_{m=1}^L \psi_m$, the measurement operators $\varphi_m=T^{-\frac{1}{2}}\psi_m T^{-\frac{1}{2}}$ are guaranteed to be orthogonal. All this sums up to the fact that the $L$ vectors $\ket{\varphi_m}=T^{-\frac{1}{2}}\ket{\psi_m}$ are orthonormal, so $\{\ket{\varphi_m}\}_{m=1}^L$ is an orthonormal basis of $S$. It only remains to calculate the performance of the transmission code that uses the states $\varphi_m$ instead of $\psi_m$ as code words, together with the unmodified decoder. The mean probability of error is given by
\begin{equation}\label{eq:perr_trdist}
  \frac{1}{L}\sum_{m=1}^L(1-\Tr\cN(\varphi_m)D_m)  \leq  \frac{1}{L}\sum_{m=1}^L \left[1-\Tr\cN(\psi_m)D_m +\frac{1}{2}\|\psi_m-\varphi_m\|_1\right]\!.
\end{equation}
We can calculate the trace distance by noticing that the fidelity $F(\psi_m,\varphi_m)^2=\Tr\varphi_m\psi_m$ satisfies
\[
\frac{1}{L}\sum_{m=1}^L \Tr\psi_m\varphi_m\geq \left(1-\frac{\varepsilon}{(1-\varepsilon)^2}\right)^2\geq1-\frac{2\varepsilon}{(1-\varepsilon)^2},
\]
and now we just use the Fuchs-van de Graaf inequality $\sqrt{1-F(\psi_m,\varphi_m)^2} \geq \frac{1}{2}\|\psi_m-\varphi_m\|_1$ \cite{FVdG:ineq} (in this case the inequality actually is an identity as the states we are comparing are both pure) to find
\[
\frac{1}{L}\sum_{m=1}^L\frac12\|\psi_m-\varphi_m\|_1\leq\sqrt{\frac{2\varepsilon}{(1-\varepsilon)^2}}.
\]
By inserting this result and Equation \eqref{eq:li_code} into Equation \eqref{eq:perr_trdist} we find the mean probability of error for the new code: 
\begin{equation}\label{eq:mean_error}
  \frac{1}{L}\sum_{m=1}^L(1-\Tr\cN(\varphi_m)D_m) \leq \frac{\varepsilon}{(1-\varepsilon)^2}+\frac{\sqrt{2\varepsilon}}{1-\varepsilon} \leq\frac{\sqrt{5\varepsilon}}{(1-\varepsilon)^2}.
\end{equation}
This is our desired code with a bound on its average error probability. Now, with the standard expurgation steps we can find a code that is half the size and has a maximum probability of error given by the double of the average probability of error. In other words, we can construct a code of size $M'=\frac{L}{2}\geq \frac{M}{2}(1-\varepsilon)^2$ and maximum probability of error bounded by:
\[
\forall m\in\cM',\quad 1-\Tr\cN(\varphi_m)D_m\leq\frac{2\sqrt{5\varepsilon}}{(1-\varepsilon)^2}=:\delta,
\]
concluding the proof of Theorem \ref{th:PureOrthogonalCode}. \qed
\end{proof}

From a transmission code, we have constructed another one with pure and orthogonal states without disrupting too much the size of the original or the error: $\delta$ is a universal function of $\varepsilon$, and $M'$ is at least a constant fraction of $M$, meaning we lose only a constant fraction of code size, without affecting the double exponential rate for large values of $n$.

Our objective now is to construct an identification code with the same scheme we have seen in the introduction: concatenating the pure and orthogonal-state transmission code with Proposition \ref{prop:A&D}. As discussed at the beginning of this section, we will replace the uniform mixture of code words from the transmission code with uniform superpositions of pure and mutually orthogonal states. This results in a double exponential set of $N_\text{sim,0}$ pure states that we can use for identification maintaining L\"ober's decoder. 

\begin{mytheorem}
\label{th:PureStateIDcode}
From an $(n,\delta)$-code for transmission consisting of pure and orthogonal code states, one can construct a zero-entropy and simultaneous identification code that achieves the same rate as L\"ober's simultaneous construction, while the errors of first and second are arbitrarily small as functions of $\delta$.
In particular, $C_\text{ID}^{\text{sim},0}(\cN)\geq C(\cN)$.
\end{mytheorem}

Note that this generalises Buhrman \emph{et al.}'s quantum fingerprinting \cite{BCW:fingerprinting}; specifically, the following proof generalises the fingerprinting construction from \cite[Proposition~5]{Winter:QCidentification}.

\begin{proof}
We start with an arbitrary $(n,\epsilon)$-transmission code with $M = 2^{n(C(\cN)-\epsilon)}$ messages, which exists for sufficiently large block length $n$. By Theorem \ref{th:PureOrthogonalCode} we can modify it to another $(n,\delta)$-code $\{(\varphi_m,D_m):m\in\cM'\}$ with the $\varphi_m=\ketbra{\varphi_m}{\varphi_m}$ pure and mutually orthogonal states, $M'=|\cM'| \geq \Omega(M)$ and $\delta=\delta(\epsilon)$. By Proposition \ref{prop:A&D} there exists a number $N\geq 2^{\lfloor\mu M'\rfloor}/M'$ of subsets $\cM_j\subset \cM'$, each of cardinality $L := \lfloor\mu M'\rfloor$ and pairwise overlap bounded by $\delta\lfloor\mu M'\rfloor$, where $\mu>0$ is a parameter small enough to fulfill $\delta\log\left(\frac{1}{\mu}-1\right)\geq1$.

The main idea now is that, instead of using the mixture of the states in each subset as code words (L\"ober's construction \cite{loeber:PhDThesis}), we use their superposition. Concretely, we define our pure code state vectors
\[
  \ket{\phi_j} := \frac{1}{\sqrt{L}} \sum_{m\in\cM_j} e^{i\alpha_{mj}}\ket{\varphi_m}, 
\]
i.e.~the states are $\phi_j = \frac{1}{L}\sum_{m,m'\in\cM_j} e^{i(\alpha_{mj}-\alpha_{m'j})}\ketbra{\varphi_m}{\varphi_{m'}}$, for all $j\in[N]$, and the simultaneous decoding operators $E_j=\sum_{m\in\cM_j}D_m$. Here, the phase angles $\alpha_{mj}\in[0,2\pi)$ are independent and uniformly random, and we shall prove that with high probability they result in all error probabilities of first and second kind being small. 
(In Appendix \ref{app:counterexample} we show by an explicit example why the phases are necessary, in the sense that the construction can fail if we fix the phases arbitrarily.)
To make the following equations more readable let us define a POVM with elements $\widetilde{D}_m = \cN^{\dagger\ox n}(D_m)$ acting directly on the input pure states, such that for all $m\in\cM'$, $\Tr\varphi_m \widetilde{D}_m \geq 1-\delta$. Likewise, we introduce $\widetilde{E}_j = \cN^{\dagger\ox n}(E_j) = \sum_{m\in\cM_j} \widetilde{D}_m$.

The first and basic observation about this construction is that the average state $\EE \phi_j$ over all the possible phases is equal to the mixture proposed by L\"ober in his construction [see Equation \eqref{eq:LoeberMixture}]. Indeed,
\[\begin{split} 
  \EE\phi_j 
    &= \int_0^{2\pi}\cdots\int_0^{2\pi} \prod_{m\in\cM_j} \frac{{\rm d}\alpha_{mj}}{2\pi} \sum_{m,m'\in\cM_j} \frac{1}{L} e^{i(\alpha_{mj}-\alpha_{m'j})} \ketbra{\varphi_m}{\varphi_{m'}}  \\
    &= \frac{1}{(2\pi)^L} \int_0^{2\pi}\cdots\int_0^{2\pi} \prod_{m\in\cM_j} \frac{{\rm d}\alpha_{mj}}{2\pi} \sum_{m\in\cM_j} \frac{1}{L} \ketbra{\varphi_m}{\varphi_{m}} 
     = \frac{1}{L}\sum_{m\in\cM_j} \varphi_m 
     =: \rho_j,
\end{split}\]
as the integral $\int_0^{2\pi}e^{i(\alpha_{m'j}-\alpha_{mj})}{\rm d}\alpha_{mj} = 0$, unless $m=m'$ in which case it results trivially in $\int_0^{2\pi}{\rm d}\alpha_{mj} = 2\pi$.

By L\"ober's construction [see Equation \eqref{eq:Lober_ErrorAnalysis}], we can assume that 
\begin{align}
  \forall j 
    \quad \EE\Tr\phi_j(\1-\widetilde{E}_j) = 1-\Tr\rho_j\widetilde{E}_j \leq \delta &=: \lambda_1, \\
  \forall k\neq j 
    \quad \EE\Tr\phi_j\widetilde{E}_k = \Tr\rho_j\widetilde{E}_k \leq \delta+\delta &=: \lambda_2.
\end{align}
For the moment we fix $j$ and look at the random variables $X_j = \Tr\phi_j(\1-\widetilde{E}_j)$ and $X_k = \Tr\phi_j\widetilde{E}_k$ (for $k\neq j$ in $\cM_j$), which are jointly distributed thanks to the uniform and independent $\alpha_{mj}\in[0,2\pi)$. Note that the above bounds imply, via Markov's inequality, that the median of $X_j$ is $\leq 2\delta$ and the median of each $X_k$ ($k\neq j$) is $\leq 4\delta$. We want to study the concentration of measure of our random variables $X_j$ and $X_k$. What we want to understand hence is the so-called \textit{concentration function} that tells us how the probability distribution behaves with respect to deviations from its median value.

To do this rigorously, we have to define the probability measure space on which we want to study the measure concentration as well as a convenient metric on it. 
Notice that our random variables $X_j$ and $X_k$ are functions of the independently uniformly distributed phase angles $\alpha_{mj}$ in $L$ orthogonal subspaces (the $L$ orthogonal states from the superposition). These phase vectors $\vec{\alpha}$ form an $L$-dimensional torus $\mathbb{T}^L$. We finally equip this space with the weighted $\ell^1$-metric
\begin{equation}
    d(\vec{\alpha},\vec{\beta}) = \frac{1}{2\pi L}\sum_{m\in\cM_j} |\alpha_{mj}-\beta_{mj}|.
\end{equation}
This $(\mathbb{T}^L,d)$ metric allows us to define the $r$-neighbourhood $A^{(r)}$ of a subset $A\subset\mathbb{T}^L$ as the set of all points in the torus that are at a distance at most $r$ from some point in $A$. Formally, $d(x,A) := \inf_{y\in A} d(x,y)$, and $A^{(r)} := \{x\in X:d(x,A) \leq r\}$. The concentration function $a(r)$ is the worst-case measure of the complement of the $r$-neighbourhood of any set of measures with normalized probability at least $\xi(A)\geq 1/2$:
\[
  a(r) := \sup_A \left\{1-\xi\!\left(A^{(r)}\right)\right\} \text{ s.t. }  \xi(A)\geq\frac12, 
\]
where $\xi$ denotes the uniform distribution on $\mathbb{T}^L$. 
For this metric space, it has been calculated in \cite[Theorem 4.4]{Ledoux-measure-concentration} and its subsequent discussion, namely $a(r) \leq e^{-r^2 L/8}$. This gives us the proportion of instances of the random phase vector that are at a distance bigger than $r$ from a given ``large'' set, which crucially goes to zero exponentially in $L$. 

Next we define the super-median sets $A_j=\{\vec{\alpha}: X_j \leq 2\delta\}$ and similarly $A_k=\{\vec{\alpha}: X_k \leq4\delta\}$, and notice that by the above Markov inequality reasoning, their measures are $\xi(A_j)\geq 1/2$. 

To apply this measurement concentration knowledge, we just need to relate the (trace) distance between two of our random superpositions with the weighted $\ell^1$-metric that we have defined. First notice that for pure states,
\[
  \frac12 \|\phi_j(\vec{\alpha})-\phi_j(\vec{\beta})\|_1 = \sqrt{1-\left|\braket{\phi_j(\vec{\alpha})}{\phi_j(\vec{\beta})}\right|^2}
    \leq \left| \ket{\phi_j(\vec{\alpha})}-\ket{\phi_j(\vec{\beta})} \right|.
\]
Now, an elementary calculation, cf.~\cite[Appendix C]{LPSW}, shows that 
\[
\begin{split}
    \left| \ket{\phi_j(\vec{\alpha})}-\ket{\phi_j(\vec{\beta})} \right|^2&=\frac{1}{L}\sum_{m\in\cM_j}|e^{i\alpha_{mj}}-e^{i\alpha_{mj}}|^2\\
    &\leq\frac{2}{L}\sum_{m\in\cM_j}|\alpha_{mj}-\beta_{mj}|=4\pi d(\vec{\alpha},\vec{\beta}).
\end{split}
\]
Since $0\leq\widetilde{E}_k\leq \1$, this means that changing the value of any $X$ by more than $\delta$, say $|X(\vec{\alpha})-X(\vec{\beta})| \geq \delta$, requires $d(\vec{\alpha},\vec{\beta}) \geq \frac{\delta^2}{4\pi}$. Putting it all together, we get strong concentration bounds of the pure state traces:
\begin{align}
  \forall j 
    \quad \Pr\left\{ \Tr\phi_j(\1-\widetilde{E}_j) > 3\delta \right\} 
    &\leq e^{-\delta^4L/128\pi^2}, \\
  \forall k\neq j 
    \quad \Pr\left\{ \Tr\phi_j\widetilde{E}_k > 5\delta \right\} 
    &\leq e^{-\delta^4L/128\pi^2}.
\end{align}
Thus, for $j\in[N']$, with $N' = \min\left\{ N, \left\lfloor e^{\delta^4L/128\pi^2} \right\rfloor-1 \right\}$, we are guaranteed by the union bound that for each $j$ there is a vector $\vec{\alpha}_j$ of phase angles such that 
\begin{align}
  \forall j\quad \Tr\phi_j(\1-\widetilde{E}_j) 
    &\leq 3\delta =: \lambda_1', \\
  \forall k\neq j 
    \quad \Tr\phi_j\widetilde{E}_k 
    &\leq 5\delta =: \lambda_2'.
\end{align}

The achieved size of the code has a double exponential growth in the block length:
\begin{equation}\label{eq:N_geq_pure}
    N' \geq 2^{\Omega\left( 2^{n(C(\cN)-\epsilon)} \right)}, 
\end{equation}
and since we can make $\epsilon$, and hence $\delta$ and the $\lambda_i'$, arbitrarily small, this proves the claim.\qed
\end{proof}

Theorem \ref{th:PureStateIDcode} proves that zero-entropy codes have double exponential growth in the block length. This poses a new fundamental difference with respect to deterministic identification where the performance is limited to exponential or slightly super-exponential.

\section{Simultaneous identification capacity is attained with pure state encodings}
\label{sec:C_train}
In the current section, we find the main result of the present work: the two right-hand capacities in Equation \eqref{eq:2D} are actually equal for all channels. 

\begin{mytheorem}
    \label{thm:sim=sim-zero}
    For every quantum channel $\cN:A\rightarrow B$, the capacity of simultaneous identification is equal to the zero-entropy capacity of simultaneous identification: 
    $C_\text{ID}^\text{sim}(\cN) =C_\text{ID}^\text{sim,0}(\cN)$.
\end{mytheorem}
\begin{proof}
We prove this result using quantum soft-covering \cite[Theorem 3.3]{TSW:soft-covering} and the ideas from the proof of \cite[Theorem 6.3]{TSW:soft-covering}. Let us start with a simultaneous $(n,\lambda_1,\lambda_2)$-ID code $\{(\rho_j,E_j) : j\in[N]\}$ for our channel $\cN$ and assuming $\lambda_1+\lambda_2<1$. As the code is simultaneous there is a measurement $(M_y)_{y\in\mathcal{Y}}$ with $\mathcal{Y}$ the set of classical outcomes such that each $(E_j,\1-E_j)$ is obtained as a coarse-graining of it. Then we can define the qc-channel $\cM:A^n\rightarrow Y$ corresponding to the measurement $\cM(\rho)=\sum_y\Tr \cN^{\otimes n}(\rho) M_y \ketbra{y}{y}$. Notice that the initial code is necessarily also an ID code for $\cM$. By applying now \cite[Theorem~3.3]{TSW:soft-covering} 
we find that for each $j$ there exists a state $\sigma_j\in \cS(A^n)$ of limited rank $r$ such that
\begin{equation}\label{eq:covering}
\frac{1}{2}\|\cM(\rho_j)-\cM(\sigma_j)\|_1\leq
\epsilon\text{\quad and\quad}\log r\leq[-H_\text{min}^\delta(A_R^n|Y)_\omega-2\log\eta]^+,
\end{equation}
where $\omega^{A_R^nY}=\sum_y\Tr_{A^n}\left(\left(\mathds{1}\otimes M_y\right)\Phi_j^{A_R^nA^n}\right)\otimes\ketbra{y}{y}^Y$ with $A_R$ the reference (isomorphic) space of $A$, $\Phi_j^{A_R^nA^n}$ is a purification of $\rho_j$, and $\epsilon=4(\delta+\eta)$ where $\delta,\eta\in(0,1)$ can be chosen freely. Furthermore, $H_{\min}^\delta(A|B)$ is the smooth min-entropy (cf.~\cite{tomamichel:book}), and $[x]^+ = \max\{x,0\}$. Notice that as $\omega^{A_R^nY}$ is a cq-state, the smooth min-entropy has to be $H_\text{min}^\delta(A_R^n|Y)_\omega\geq 0$, therefore $r\leq 1/\eta^2$.

From Equation \eqref{eq:covering} it is clear that we can use the states $\{\sigma_j\}_{j\in[N]}$ to create a simultaneous $(n,\lambda_1+\epsilon,\lambda_2+\epsilon)$-ID code for our channel. Furthermore, as these have a limited rank $r$ we can purify them into $\ket{\zeta_j}\in A^n\otimes A^\ell$ as long as $|A|^\ell \geq 1/\eta^2 \geq r$ (that is: $\ell=\lceil\log r/\log |A|\rceil$), and send the pure code words through $n+\ell$ uses of the channel. Using the trivial extension of the decoding operators $E_j\otimes\1_B^{\otimes\ell}$, corresponding to first tracing out $A^\ell$ and then using the previous decoding effect $E_j$ on $A^n$, we end up with a zero-entropy simultaneous $(n+\ell,\lambda_1+\epsilon,\lambda_2+\epsilon)$-ID code. 

The proposed construction will obviously make the rates a little smaller, as we have the same number of messages over a larger block length. 
However, $\ell \leq O(\log r) \leq O(\log\frac{1}{\eta})$, which is a constant independent of $n$. Thus,
$n \sim n+\ell$ and we get for the rate of the new $(n+\ell,\lambda_1+\epsilon,\lambda_2+\epsilon)$-ID code $\{(\zeta_j,E_j\ox\1_B^{\ox\ell}):j\in[N]\}$: 
\[
  \liminf_{n\rightarrow\infty} \frac{1}{n+\ell} \log\log N 
  = \liminf_{n\rightarrow\infty} \frac{1}{n} \log\log N.
\]
In conclusion, from a simultaneous ID code, we can create a zero-entropy and simultaneous ID code with only slightly larger error probabilities and the same asymptotic rate, meaning that $C_\text{ID}^\text{sim,0}(\cN) \geq C_\text{ID}^\text{sim}(\cN)$.

In Section \ref{sec:pure-encoders} we have commented on the opposite inequality $C_\text{ID}^{\text{sim},0}(\cN)\leq C_\text{ID}^\text{sim}(\cN)$, which trivially emerges from the fact that using zero-entropy encoders impose a further restriction on simultaneous identification codes. This completes the proof.\qed
\end{proof}

We remark that this yields a second, independent proof of Theorem \ref{th:PureStateIDcode}, since L\"ober had already shown double exponential growth of the message set under simultaneous identification, and indeed $C_\text{ID}^\text{sim}(\cN) \geq C(\cN)$; now Theorem \ref{thm:sim=sim-zero} provides the conversion of this simultaneous ID code into one with a zero-entropy encoder and the same rate. However, as we have a very direct proof of Theorem \ref{th:PureStateIDcode} above, along the lines of Ahlswede/Dueck and L\"ober, we feel that it makes sense to state them both. In any case, this allows us to collapse the two-dimensional set of relations in Equation \eqref{eq:2D} into the following chain of inequalities:
\begin{equation}
\label{eq:train_capacities}
    C_\text{ID}(\cN) 
     \geq C_\text{ID}^0(\cN) 
     \geq C_\text{ID}^\text{sim}(\cN)
     =    C_\text{ID}^\text{sim,0}(\cN)
     \geq C(\cN),
\end{equation}
answering, in particular, the question posed in the introduction regarding the relation between $C_\text{ID}^0$ and $C_\text{ID}^\text{sim}$.
In the next section, we are going to discuss some simple examples showing that both the leftmost and the middle inequality can be either strict or equal. It remains an open question, as it had been since \cite{loeber:PhDThesis}, whether the rightmost inequality is actually an identity. 
No separation between the simultaneous ID capacity and the classical transmission capacity is known, although possible candidate channels to look for one are identified in certain qc-channels in \cite{Winter:Review}. There is no compelling reason to think that the two capacities should always be the same.

\section{Separations between capacities and further analysis}
\label{sec:examples}
\begin{myexample}\label{example:id}
    The unrestricted, simultaneous and zero-entropy capacities of identification for the ideal (noiseless) channel $\id_A$, for a general Hilbert space $A$ of dimension $|A|$ satisfy
    \begin{equation}
       2\log|A|= C_\text{ID}(\id_A)>C_\text{ID}^0(\id_A)=C_\text{ID}^\text{sim}(\id_A)=\log|A|.
    \end{equation}
\end{myexample}
\begin{proof}
We know from \cite{Winter:QCidentification} that $C_\text{ID}(\id_A)=2\log|A|$ and from the recent preprint \cite{TSW:soft-covering} that $C_\text{ID}^\text{sim}(\id_A)=\log|A|$. We now invoke an auxiliary result from \cite{Winter:QCidentification,HLW:general_entanglement}:
\begin{myproposition}
\label{prop:upperbound_sizeN}
Let $\{(\rho_j,D_j):j\in[N]\}$ be an ID code on a space of dimension $d$ and error probabilities $\lambda_1$, $\lambda_2$ of first and second kind, respectively, with  $\lambda_1+\lambda_2<1$. Then:
\begin{equation*}
        \begin{split}
            \text{if all }\rho_j\text{ are pure,}\quad & N\leq\left(\frac{5}{1-\lambda_1-\lambda_2}\right)^{2d},\\
            \text{while for general }\rho_j,\quad & N\leq\left(\frac{5}{1-\lambda_1-\lambda_2}\right)^{2d^2}.
        \end{split}
\end{equation*}
\end{myproposition}
With this, we can prove that the ID capacity of the identity channel using zero-entropy encoders is $C_\text{ID}^0(\id_A) =\lim_{n\rightarrow\infty}\frac{1}{n}\log\log N_0(n,\lambda_1,\lambda_2)=\log|A|$. Indeed, the above upper bound concerning pure states with $d=|A|^n$ on the one hand, and on the other the lower bound from Theorem \ref{th:PureStateIDcode} [Equation \eqref{eq:N_geq_pure}] with $C(\id_A)=\log|A|$,  together clearly imply $C_\text{ID}^0(\id_A)=\log|A|$. 

In fact, the arguments for all three capacities show that they obey the strong converse for all error probabilities $\lambda_1,\lambda_2>0$ with $\lambda_1+\lambda_2<1$.\qed
\end{proof}

We can generalise this example slightly so as to separate all three of the above identification capacities. 

\begin{myexample}
\label{example:id_x_Tr}
The zero-entropy capacity of identification for trivially extended noiseless channel $\id_A\otimes\Tr_C: \cS(A\cross C)\rightarrow\cS(A)$ (i.e. a perfect channel on $A$ together with a trace $\Tr_C:\cS(C)\rightarrow\{0\}$ over another space $C$) satisfies
\begin{equation}
  \label{eq:C-ID-0:examples}
  C_{\text{ID}}^0(\id_A\otimes\Tr_C) 
      = \begin{cases}
          \log|A| + \log|C| & \text{if } 1\leq|C|\leq|A|,\\
          2\log|A| & \text{if } |C|\geq|A|.
        \end{cases}
\end{equation}
This means that $C_{\text{ID}}^0$ can be equal to $C_\text{ID}$ or equal to $C_{\text{ID}}^\text{sim}$, but can also take any logarithm of an integer offset value between the unrestricted and the simultaneous identification capacity, depending on the dimension of the trivial extension:
\begin{equation}
       2\log|A| 
         = C_\text{ID}(\id_A\otimes\Tr_C)
         \geq C_\text{ID}^0(\id_A\otimes\Tr_C)
         \geq C_\text{ID}^\text{sim}(\id_A\otimes\Tr_C)
         = \log|A|.
\end{equation}
Furthermore, the strong converse holds for all three identification capacities, for all error probabilities $\lambda_1,\lambda_2>0$ with $\lambda_1+\lambda_2<1$. 
\end{myexample}
\begin{proof}
It is clear from their respective definitions that $C_{\text{ID}}$ and $C_{\text{ID}}^{\text{sim}}$ are unchanged by the tensoring with the partial trace channel: 
\begin{align}
  C_{\text{ID}}(\id_A\otimes\Tr_C) &= C_{\text{ID}}(\id_A) = 2\log|A|, \\
  C_{\text{ID}}^{\text{sim}}(\id_A\otimes\Tr_C) &= C_{\text{ID}}^{\text{sim}}(\id_A) = \log|A|.
\end{align}
This is because the code properties only depend on the output system $A$, and we can replace all signal states $\rho_j$ by $\rho_j' = (\Tr_{C^n}\rho_j)\otimes\sigma_0^{C^n}$, with a constant state $\sigma_0$ on $C^n$. 

On the other hand, the zero-entropy ID capacity varies with the dimension of the auxiliary system $C$. As intuition for this fact, consider an unrestricted ID code for $\id_A$. We have seen that the capacity in this case is $C_\text{ID}(\id_A)=2\log|A|$. Now we purify all the code words using an auxiliary system of the same dimension $|A|^n$ for block length $n$ and send them through a channel $\id_A\otimes \Tr_A$ that immediately discards the purifying extension. Our code book consists only of pure states, but the output is exactly the same as the initial ID code for $\id_A$, therefore the capacity has to be the same $C_\text{ID}^0(\id_A\otimes \Tr_A)=C_\text{ID}(\id_A)=2\log|A|$. On the other hand, we have seen in the previous example that $C_\text{ID}^0(\id_A)=\log|A|$. 

To prove the capacity formula \eqref{eq:C-ID-0:examples} in general, we start with the converse (upper bound). 
For this purpose, consider a given zero-entropy ID block code $\{(\psi_j,D_j):j\in[N]\}$ with pure states $\ket{\psi_j}\in A^nC^n$. Since the decoding operators only concern $A^n$, we get a code with exactly the same parameters when replacing every $\psi_j$ by an arbitrary purification of $\Tr_{C^n}\psi_j$, i.e.~$\Tr_{C^n}\psi_j' = \Tr_{C^n}\psi_j$. Thus, for $|C|>|A|$ we can modify the original ID code without loss by the purified code $\{(\psi_j',D_J):j\in[N]\}$
with $\ket{\psi_j'}\in A^nC^{\prime n}$, where $C'\subset C$ is a subspace of dimension $|C'|=|A|$. Therefore, we may from now on assume w.l.o.g.~$|C|\leq|A|$. We can then invoke once more Proposition \ref{prop:upperbound_sizeN} to obtain that for this purified code with code words in dimension $d=|A|^n|C|^n$,
\[
  N\leq \left(\frac{5}{1-\lambda_1-\lambda_2}\right)^{2|A|^n|C|^n}.
\] 
Hence, for any fixed error probabilities such that $\lambda_1+\lambda_2<1$, we find the strong converse property 
\[
  \limsup_{n\rightarrow\infty} \frac1n \log\log N 
  \leq \log|A|+\log|C|.
\]

For the achievability part (lower bound), we basically adapt the proofs of \cite[Propositions~6~and~17]{Winter:QCidentification}. We take a zero-entropy ID code in $\CC^S\subset\CC^d\otimes\CC^{d'}$ with code words $\ket{\psi_j}$, where $d\geq d'\geq\Omega(d)$, and perform the partial trace over the prime system to find $\Tr_{\CC^{d'}}\psi_j\in \mathcal{S}(\CC^d)$. These marginal states define an ID code on $\CC^d$ of size $N\geq\exp(\lfloor\epsilon K(\lambda)d^2\rfloor)/ K(\lambda)d^2$, with $\epsilon, K(\lambda)>0$.

Now, we can start with a zero-entropy (fingerprinting \cite{BCW:fingerprinting}) code in the Hilbert space $A^nC^n$ (note that we only have to deal with $|C|\leq|A|$), and consider $d=|A|^q$ and $d' \leq |A|^{n-q}|C|^n$ such that indeed $\CC^d\otimes\CC^{d'} \subset A^nC^n$. Notice that our channel applied $n$ times, $\cN^{\otimes n}=(\id_A\otimes\Tr_C)^{\otimes n}$, automatically discards the $C^n$ part of the initial system, and our code requires the receiver to discard the remaining $A^{n-q}$. This construction is equivalent to the one in the previous paragraph, meaning that we get a code of $N$ messages. Finally, as $d\geq d'$ we find the relation $|A|^q\geq |A|^{n-q}|C|^n$ as necessary and sufficient to build the code, or equivalently $2q\log|A|\geq n(\log|A|+\log|C|)$.
This directly implies that 
\[\liminf_{n\rightarrow\infty} \frac1n \log\log N \geq \log|A|+\log|C|,\]
completing the proof.\qed
\end{proof}

\medskip
Finally, we can add to the chain of inequalities \eqref{eq:train_capacities} the existing results for identification using only separable states at the encoder. In this scenario, we obtain double exponential capacities of simultaneous \cite{loeber:PhDThesis} and non-simultaneous \cite{AW:StrongConverse} identification equal to the (single exponential) capacity of transmission of the channel with separable states encoding, which is equal to the channel's Holevo capacity $\chi(\cN)$ \cite{holevo:capacity,SW:capacity}:
\(
  C_\text{ID}^\text{sep}(\cN)=C_\text{ID}^\text{sep,sim}(\cN)=C^\text{sep}(\cN)=\chi(\cN).
\)

These results are usually expressed as regarding the identification capacities of classical-quantum (cq-)channels. Indeed, without loss of generality, all input states of a cq-channel are separable; vice versa, restricting the encodings to separable states effectively reduces the channel $\cN:A\rightarrow B$ to the cq-channel $W:\mathcal{S}(A)=:\mathcal{X} \rightarrow B$ (in the sense that $\cN$ and $W$ have the same single-copy and block output states).
Notice further that, by definition, $C_\text{ID}^\text{sim}(\cN)\geq C_\text{ID}^\text{sep,sim}(\cN)$, and therefore we can extend our chain of double exponential capacity inequalities as follows:
\begin{equation}
  \label{eq:new_train}
  C_\text{ID}(\cN)\geq C_\text{ID}^0(\cN)\geq C_\text{ID}^\text{sim}(\cN)\geq C_\text{ID}^\text{sep}(\cN).
\end{equation}

\section{Conclusions}
\label{sec:conclusions}
Starting from the intuition of deterministic identification via classical channels, we have found a capacity that interpolates between the unrestricted identification capacity and the simultaneous identification capacity for any channel $\cN$, by imposing the restriction of zero-entropy (pure state) encodings.

However, the final results clearly indicate that zero-entropy encoders do not capture the essence of deterministic identification, as the rates we achieve are double exponential in the block length, a performance that cannot be achieved by classical deterministic identification, which is limited to exponential or slightly super-exponential behaviour. There is another feature of the classical deterministic identification, which lends itself to quantum generalisation, namely that the inputs of a deterministic ID code are product distributions. In concurrent work \cite{CDBW:ID-continuous-discrete}, we analyze the analogous quantum channel constraint, namely that the encoding states have to be product states (in addition to a possible further arbitrary restriction on the local states), and show that slightly super-exponential scaling of the code size ensues generally.

These new insights into the relation between identification using different encoders and decoders can be related to Freeman Dyson's analysis regarding the design of experiments, measurements, and detectors in particle physics \cite{Dyson:personal}. Notice that if we analyse the process from an operational perspective, tailoring an experiment is similar to devising clever ways to encode and decode information. Current practices in particle physics often centre on highly specific questions with binary answers, such as the detection of a particular particle following a collision. Detectors are, in these cases, optimized not to categorize every received particle, but solely to detect the specific one that researchers seek. From an operational standpoint, the distinction between these particle sensing tasks closely resembles the transmission versus identification (and, in particular, non-simultaneous identification) decoding discussion presented in the introduction.

Freeman Dyson challenged the prevailing paradigm in particle physics experimentation by advocating more universal measurement techniques that enable the extraction of relevant information through classical post-processing. From an operational viewpoint, this is similar to simultaneous identification, where the measurements are not message-dependent. Simultaneous and non-simultaneous identification achieve similar (double exponential) performance levels, reinforcing Freeman Dyson's intuition that more universal detectors, coupled with post-processing algorithms, could be devised in particle physics without compromising much the quality or certainty of outcomes.

Nonetheless, a significant challenge remains, as the encoding process in particle physics is often beyond the control of experimentalists and determined by nature itself. Even more, it is sometimes this very ``encoding'' that we want to understand when performing the experiments. The analysis of identification with different encodings (such as the present zero-entropy encoder) can provide further insight into Dyson's conjectures.

\begin{credits}
\subsubsection{\ackname} 
The authors thank Alan Sheretz and Adam Beckenbaugh for invaluable insights into erroneous message identification under various constraints, dating back to several discussions with the last author in Benson, AZ. 
H.B. and C.D. are supported by the German Federal Ministry of Education and Research (BMBF) within the national initiative on 6G Communication Systems through the research hub 6G-life under grant 16KISK002, within the national initiative on Post Shannon Communication (NewCom) under Grant 16KIS1003K, within the national initiative QuaPhySI -- Quantum Physical Layer Service Integration under grant 16KISQ1598K and within the national initiative ``QTOK -- Quantum tokens for secure authentication in theory and practice'' under grant 16KISQ038. They have also been supported by the initiative 6GQT (financed by the Bavarian State Ministry of Economic Affairs, Regional Development and Energy). H.B. has further received funding from the German Research Foundation (DFG) within Germany’s Excellence Strategy EXC-2092 -- 390781972.  
A.W. is supported by the European Commission QuantERA grant ExTRaQT (Spanish MICIN project PCI2022-132965), by the Spanish MICIN (projects PID2019-107609GB-I00 and and PID2022-141283NB-I00) with the support of FEDER funds, by the Spanish MICIN with funding from European Union NextGenerationEU (PRTR-C17.I1) and the Generalitat de Catalunya, and by the Alexander von Humboldt Foundation. P.C. and A.W. are furthermore supported by the Institute for Advanced Study of the Technical University Munich.

\vspace{-0.2cm}
\subsubsection{\discintname}
The authors declare no competing interests. 
\end{credits}

\appendix
\section{Fixed superposition phases fail to prove Theorem \ref{th:PureStateIDcode}}
\label{app:counterexample}
Using random phases in the code words for our zero-entropy ID code can seem counterintuitive. In L\"ober's construnction of a simultaneous ID code as mixtures \cite{loeber:PhDThesis}, random phases are not necessary. Our construction follows similar arguments, but replaces the mixtures by superpositions. Here we show an example that illustrates why arbitrarily fixing the phases $e^{i\alpha_{jm}}$ of our superposition fails to make the error of first kind small. As the scalars can be absorbed into the basis states $\ket{\varphi_m}$, it is enough to consider the case $\alpha_{jm}=0$.


Let us assume an orthonormal basis $\{\ket{\varphi_m}\}_{m\in\cM}$ and a POVM $(F_m:m\in\cM)$ such that $\Tr \varphi_m F_m \geq 1-\delta$ for all $m$. We define a uniform superposition of the basis states of a subset $\cM_j\subset\cM$ of cardinality $\abs{\cM_j}=K$ that could form a signal state of our ID code as $\ket{\psi_j}=\frac{1}{\sqrt{K}}\sum_m^K\ket{\varphi_m}$ and also our candidate detection effect $D_j=\sum_m^K F_m$. This is the same construction as in Theorem \ref{th:PureStateIDcode} but fixing all the phases to be (without loss of generality) equal to $1$, and for it to work we would need that $\Tr\psi_j D_j\geq 1-f(\delta)$, with $f(\delta)$ some function that decreases to $0$ with $\delta$. Although this might seem very natural to hold (especially in light of the straightforward arguments in \cite{loeber:PhDThesis}), we prove that $\Tr\psi_j D_j$ can be made arbitrarily small, in fact zero, for a suitably unfortunate set of transmission decoders. In the sequel, we omit the subscript $j$ that keeps track of the particular subset we are looking at as the argument holds equally for all of them.

Consider the following POVM: for $m=1,\dots,K$ define $F_m=\ketbra{\nu_m}{\nu_m}$ with
$\ket{\nu_m}=\ket{\varphi_m}-\frac{1}{\sqrt{K}}\ket{\psi}$
a set of $K$ subnormalized states all orthogonal to the superposition $\ket{\psi}$. For $m=K+1,\dots,M-1$, define $F_m=\ketbra{\varphi_m}{\varphi_m}$, and finally $F_M=\ketbra{\varphi_M}{\varphi_M}+\ketbra{\psi}{\psi}$. Notice that all the elements are positive semidefinite operators that sum up to the identity so they indeed form a POVM. Notice furthermore that for $m=1,\ldots,K$, $\Tr\varphi_mF_m=(1-\frac{1}{K})^2$, and for $m=K+1,\dots,M$, $\Tr\varphi_mF_m=1$. So indeed, we have $\Tr \varphi_mF_m \geq 1-\delta = 1-\frac{2}{K}$ for all possible $m$, with $\delta$ arbitrarily small for large enough subsets. 

The POVM we have defined fulfils all the necessary criteria to be used as a decoder for transmission. However, see what happens when we try to use it to create an identification code with uniform superpositions of the input transmission states. Our candidate decoding effect is
\begin{equation*}\begin{split}
  D &= \sum_{m=1}^K F_m 
     = \sum_{m=1}^K \left[\ketbra{\varphi_m}{\varphi_m}+\frac1K\ketbra{\psi}{\psi}-\frac{1}{\sqrt{K}}\ketbra{\varphi_m}{\psi}-\frac{1}{\sqrt{K}}\ketbra{\psi}{\varphi_m}\right]\\
    &= \left(\sum_{m=1}^K \ketbra{\varphi_m}{\varphi_m}\right) -\ketbra{\psi}{\psi},
\end{split}\end{equation*}
and therefore $\Tr\psi D=0$. This result implies that given a valid transmission POVM, the proposed construction with uniform superposition fails to create an ID code. Notice that the probabilistic mixture of the transmission code words (L\"ober's construction \cite{loeber:PhDThesis}) does not suffer this problem, implying that this is a pure interference effect, i.e. due solely to the coherences in our superpositions. 

The idea of using random phases to overcome this problem comes from the realization that the probabilistic mixture of a state can be seen as a dephased version of a superposition, i.e.~as the output of the channel that applies uniformly random diagonal phase unitaries to the superposition. Adding random phases to our pure state $\ket{\psi'}=\frac{1}{\sqrt{K}}\sum_{m=1}^Ke^{i\alpha_m}\ket{\varphi_m}$ results in a detection probability 
\[
\begin{split}
     \Tr\psi'D 
    &= \Tr\left[\left(\frac{1}{K}\sum_{m,m',k}^K\!\!e^{i(\alpha_m-\alpha_{m'})}\ket{\varphi_m}\!\!\braket{\varphi_{m'}}{\varphi_k}\!\!\bra{\varphi_k}\right)-\ketbra{\psi'}{\psi'}\psi\rangle\!\!\bra{\psi}\right]\\
    &=1-|\!\braket{\psi}{\psi'}\!|^2,
\end{split}
\]
and depending on the phases, the inner product between $\psi$ and $\psi'$ can indeed be small. In Theorem \ref{th:PureStateIDcode} we show that 
this is the case with high probability.

\printbibliography
\end{document}